\documentstyle[11pt,aaspp,psfig,side]{article}
\def\gs{\mathrel{\raise0.35ex\hbox{$\scriptstyle >$}\kern-0.6em % Greater/squiggles
\lower0.40ex\hbox{{$\scriptstyle \sim$}}}}
\def\ls{\mathrel{\raise0.35ex\hbox{$\scriptstyle <$}\kern-0.6em % Less than/squiggles
\lower0.40ex\hbox{{$\scriptstyle \sim$}}}}

\def\arcsper{\ifmmode \rlap.{''}\else $\rlap{.}''$\fi}
\def\arcmper{\ifmmode \rlap.{'}\else $\rlap{.}'$\fi}

\def\RR{\hbox{$R_{702}$}}
\def\II{\hbox{$I_{814}$}}
\def\VV{\hbox{$V_{555}$}}
\def\BB{\hbox{$B_{450}$}}
\received{29-7-96}
\revised{12-11-96}
\accepted{26-11-96}

\begin{document}
%\small
%\doublespace

\title{A CATALOG OF MORPHOLOGICAL TYPES IN \\
10 DISTANT RICH CLUSTERS OF GALAXIES\footnote{
Based on observations 
obtained with the NASA/ESA Hubble Space Telescope
which is operated by STSCI for the Association of Universities
for Research in Astronomy, Inc., under NASA contract NAS5-26555.}}
\author{
Ian Smail\altaffilmark{1}\footnote{Visiting Research Associate at the Carnegie
Observatories.},
Alan Dressler\altaffilmark{2},
Warrick J.\ Couch\altaffilmark{3},\\
Richard S.\ Ellis\altaffilmark{4},
Augustus Oemler Jr.\altaffilmark{5}\footnote{Present address: 
 The Observatories of the Carnegie Institution of Washington, 813 Santa Barbara St., Pasadena, CA 91101-1292},
Harvey Butcher\altaffilmark{6} \&
Ray M.\ Sharples\altaffilmark{1}.
}

\bigskip

\affil{ 1) Department of Physics, University of Durham, South Rd, Durham DH1 3LE, UK}
\affil{ 2) The Observatories of the Carnegie Institution of Washington, 813 Santa Barbara St., 
Pasadena, CA 91101-1292}
\affil{ 3) School of Physics, University of New South Wales, Sydney 2052, Australia}
\affil{ 4) Institute of Astronomy, Madingley Rd, Cambridge CB3 OHA, UK}
\affil{ 5) Astronomy Department, Yale University, PO Box 208101, New Haven CT 06520-8101}
\affil{ 6) NFRA, PO Box 2, NL-7990, AA Dwingeloo, The Netherlands}
%\vfil\eject

\begin{abstract}
We present catalogs of objects detected in deep images of 11 fields in
10 distant clusters obtained  using WFPC-2 on board the {\it Hubble
Space Telescope}.  The clusters span the redshift range
$z=0.37$--$0.56$ and are the subject of a detailed ground- and
space-based study to investigate the evolution of galaxies as a
function of environment and epoch.  The data presented here include
positions, photometry and basic morphological information on $\sim
9000$ objects in the fields of the 10 clusters.  For a brighter subset
of 1857 objects in these areas we provide more detailed morphological
information.
\end{abstract}

\keywords{galaxies: clusters: general -- galaxies: evolution -- 
galaxies: structure}

\sluginfo
\newpage

\section{Introduction}

Except for the simplest cases, there is significant ambiguity in
inferring a galaxy's star formation history from its current integrated
properties.  However, substantial progress has been made in the study
of galaxy evolution through the use of the lookback in cosmic time
afforded by observations of distant cluster and field galaxies.
Observations of the properties of galaxies at much earlier times can
help to distinguish between the possible star formation models and thus
offer a clearer picture of the evolutionary paths taken by galaxies as
a function of morphology, mass, and environment.

One of the clearest cases of evolution in a galaxy population is that
shown by the galaxies in the cores of rich clusters.  The surprisingly
rapid bluing of this population with increasing redshift was revealed
by Butcher \& Oemler (BO, 1978, 1984).  Spectroscopic surveys of
galaxies in these distant clusters (e.g.\  Couch \& Sharples 1987;
Dressler \& Gunn 1992) uncovered further spectral signatures of
evolutionary changes, including high rates of star formation and
evidence for strong star bursts.  There has been extensive discussion
of possible mechanisms for causing the BO effect, although no
comprehensive solution has yet been proposed.  Nevertheless, it has
become apparent that the growth of clusters, with its attendant influx
of field galaxies, cannot be viewed in isolation from the changes ongoing 
in the surrounding field population (e.g.\ Broadhurst, Ellis \&
Shanks 1988; Cowie et al.\ 1996).    We are therefore undertaking an
ongoing study of the evolution of cluster and supercluster populations
at intermediate redshifts to tie the population changes in these
environments together and thus provide a unified picture of
environmental influences on the evolution of galaxies.

In this paper we present data for the initial study of galaxies in the
central regions of rich clusters at intermediate redshifts, observed
some 2--4$h^{-1}$ billion years before the present day.\footnote{We use
$q_o=0.5$ and adopt the dimensionless Hubble parameter,
$h = H_o / 100$ km sec$^{-1}$ Mpc$^{-1}$.
This geometry means that 1 arcsec is equivalent to 3.09 h$^{-1}$ kpc
for our lowest redshift cluster and  3.76 h$^{-1}$ kpc in the most
distant.} Similar studies have been made before (e.g.\ Butcher \&
Oemler 1984; Couch \& Sharples 1987; Dressler \& Gunn 1992), but the
advent of the {\it Hubble Space Telescope} (HST) with its high spatial
resolution has added a vital new dimension to the effort ---
morphological information that can be used to link the evolution of
stellar populations with changes in galaxy {\it structure}, in order to
understand how the wide range of galaxies we see today came to be.
Pre-refurbishment HST observations by two groups (Couch et al.\ 1994;
Dressler et al.\ 1994; Oemler et al.\ 1996) were used in early attempts
to correlate spectral evolution with morphological/structural data, and
to provide some insight into the mechanisms that might be driving the
strong evolution in the cluster galaxy population.  These two programs
were combined for Cycle-4 into the ``MORPHS'' project.  This program
has accumulated post-refurbishment HST images for 11 fields in 10
intermediate redshift ($0.37 < z < 0.56$) clusters.  Here we provide
the basic data derived from these images, including catalogs of
parameters for objects in the fields, and morphological information ---
both algorithm-derived parameters that provide a rudimentary description
of the galaxies, and more detailed visual classification onto the
Revised Hubble system.  The catalogs, as well as the processed images
themselves, are available in CD-ROM form.

The catalogs and imaging data, along with the ground-based spectroscopy
for galaxies in these fields, form the basis of the analysis which is
presented elsewhere.  In particular we discuss the morphology-density
relation in these distant clusters in Dressler et al.\ (1996), while
the evolution in the cluster spheroidal galaxy populations is addressed
in terms of the homogeneity of their colors by Ellis et al.\ (1996) and
through changes in their structural parameters in Barger et
al.\ (1996).    An analysis of gravitational lensing by the clusters is
given in Smail et al.\ (1996).  Other features of the galaxy
populations in these distant clusters, as well as the extensive
spectroscopic dataset we have amassed, will be dealt with in
forthcoming papers.

A plan of the paper follows. The imaging data on the 11 fields, their
reduction and cataloging are discussed in \S 2.  We present the scheme
adopted for the morphological classification of brighter subsamples in
the clusters in \S 3.  Finally, in \S 4 we give some basic statistics
about the galaxy populations in these distant clusters.

\section{Reduction and Analysis}

\subsection{Observations}
  
The lack of a large, well defined catalog of distant clusters hampers
selection of targets for studies  of galaxy evolution as a function of
lookback time and environment.  Therefore, the 10 clusters included in
this study were not chosen according to strict criteria, such as X-ray
luminosity or optical richness, but rather represent the best studied examples
at intermediate redshift, $z\sim 0.5$.  The clusters span a wide range
in X-ray luminosity (Table~1), optical richness (Fig.~5), and mass (Smail et
al.\ 1996).
 
All the clusters were imaged using the {\it Wide Field and Planetary
Camera 2} (WFPC-2) on board HST.  In total 71 primary spacecraft orbits
were allocated in Cycle-4 to image fields in 9 of these clusters.
These pointings comprised 7 core regions and 2 pointings on fields
around 3--5 arcmin from the cores in 2 clusters (A370 and
Cl0939$+$47).  Additional pointings of the centers of 2 more clusters
were obtained during the ERO phase of Cycle-4 (10 orbits on
Cl0939$+$47) and through the generosity of Turner and collaborators who
provided their imaging of Cl0024$+$16 (a further 18 orbits).  The
complete cluster sample thus comprises images of the central regions of
8 clusters, the outer regions of another cluster (A370) and both a
center and outer field on the final cluster (Cl0939$+$47).  Details of
these, together with relevant observational information (field
position, orientation, exposure times and surface brightness limits),
are given in Table~1.   We also list in Table~1 the number of objects
in the full catalog of each field, as well as in the morphological
samples.  Finally, we list some general information about the clusters,
including rest-frame velocity dispersions from Dressler \& Gunn
(1992)\footnote{These values have been corrected for errors in the
original paper resulting from the use of an incorrect sample size in
the calculation of the velocity dispersions.} and Soucail et
al.\ (1988), X-ray luminosities in the 0.3--3.5 keV band (\cite{fjc};
\cite{h82}; \cite{ws}; \cite{irs96}) and reddening.
  
The filters used for the observations discussed here are F450W (\BB),
F555W (\VV), F702W (\RR) and F814W (\II).  The individual exposures
were generally grouped in sets of 4 single-orbit exposures each offset
by 2.0 arcsec to allow for hot pixel rejection. After standard pipeline
reduction, the images were aligned using integer pixel shifts and
combined using the IRAF/STSDAS task CRREJ.  The PC chip was then
rebinned to the same linear scale as the WFC and the sky levels in the
4 chips equalized, before assembling them into a mosaic.  The mosaicing uses
integer pixel shifts to roughly position the 4 chips relative to each
other.  This is sufficient to align objects lying across the chip
boundaries at the $<2$ pixel level and has the advantage over a true
astrometric mosaic that the data are not resampled.  We chose to retain
the WFPC-2 color system and hence use the F450W, F555W, F702W and F814W
zero points from Table~9 of Holtzman et al.\ (1995).  The actual zero
point used in the photometry of each frame is written in the header of
the image on the CD-ROM under the keyword {\sc MAGZPT}.  The headers
also provide all the standard WFPC-2 pipeline reduction information.
The final images cover the central 0.4--0.8 h$^{-1}$ Mpc of the
clusters (Plates~1a-k) to a 5$\sigma$ limiting depth of $I_{814} \simeq
26.0$ or $R_{702} \simeq 27.0$.  

\subsection{Constructing Object Catalogs}

To catalog faint objects in these frames and measure their shapes we
use a modified version of the SExtractor image analysis package
(\cite{eb95} and below).  We adopt a fixed detection isophote
equivalent to $\sim 1.3 \sigma$ above the sky, where $\sigma$ is the
average standard deviation of the sky noise in the frames measured
pixel to pixel, viz.\ $\mu_{814}= 25.0$ mag arcsec$^{-2}$ ($\mu_{814} =
24.6$ mag arcsec$^{-2}$ for A370\#2 and Cl0939$+$47\#2) or $\mu_{702} =
25.7$ mag arcsec$^{-2}$ for the WFC chips, and a minimum area after
convolution with a 0.3 arcsec diameter top-hat filter of 0.12
arcsec$^{2}$.  For the lower sensitivity PC region we used a detection
isophote 1 magnitude brighter than the equivalent WFC value.   The
object catalogs are all constructed from the reddest passband available
for a given cluster (either the F702W or F814W exposures).  For those
clusters where we have observations in more than one passband we
measured the colors for cataloged objects in 1.0 arcsec diameter
matched apertures using the IRAF task PHOT.  It should be noted that we
have applied no reddening corrections to {\it any} of the photometry
published in this paper.  However, for reference we list in Table~1 the
standard reddening estimates for the fields, these indicate that the
correction is only appreciable in a single field (Cl0303$+$17).

We modified the SExtractor code to calculate two additional
parameters:  the concentration index ($C_1$), which is the fraction of
an object's light contained in the central 30\% of its area as measured
in an ellipse aligned with the object and having the same axial ratio
(c.f.\ \cite{ra95} \& Fig.~3); and the contrast index ($C_2$), which
measures the fraction of an object's light contained in the brightest
30\% of its pixels.  The concentration index is a measure of how
centrally peaked an object is, while the contrast index provides a
similar measure,  but without requiring the light to be centrally
concentrated.  A comparison of these two indices can thus be used to
determine if an object has a smooth (when the two indices have similar
values) or clumpy light distribution (when the contrast exceeds the
concentration).

After automated detection and deblending, the object catalogs for each
frame were visually inspected and spurious objects (e.g.\ diffraction
spikes) were removed.  In addition it was found that on rare
occasions the detection routine would fail to catalog a particular
galaxy, usually owing to its proximity to another brighter object.  For
completeness SExtractor was therefore re-run on small regions of the
frames centered on these objects with a more vigorous deblending
criterion and the cataloged parameters for these ``additional'' objects
added to the full catalogs.  Such objects have been flagged in the
catalogs by having IDs of 3000 and above.  The analysis of our frames
provides final catalogs of $\sim 800$ objects in each of the clusters
(Tables~1 and 3 a-k).

\section{Morphological Classifications}

\subsection{Visual Typing}

While some basic morphological information is available from parameters
in the automated catalogs, it was felt that detailed visual
classification of galaxies onto the Revised Hubble scheme is still
essential.  The uniqueness of the dataset and the high quality
information contained in it more than justified  the large investment
of time  required.  It is apparent that robust morphological
classification relies upon an object's image having sufficiently high
signal-to-noise, at the level of at least a few hundred.  Thus, not
every object in the full catalogs can be reliably classified.  We
therefore chose to limit the morphological classification to objects
brighter than a fixed apparent magnitude, although the signal-to-noise
limitation actually sets in over a range of about a magnitude, due to
the dispersion in galaxy size and surface brightness.  After some
experimentation we chose $R_{702} \leq 23.5$ or  $I_{814} \leq 23.0$ as
the limit beyond which the classifications became 
unreliable.  Here we define unreliable to mean that in more than 20\%
of cases independent classifiers differed by more than 1 class in
assigning Revised Hubble types  (including cases where classifiers
where unable to assign a type).  Typically there are $\sim 170$
galaxies per cluster (Tables~1 and 4a-k) to these limits, ranging
between 116 in the outer field of Cl0939$+$47, up to 248 for Cl0024$+$16.
These objects are identified on Plates 1a-k.

The assignment of basic morphological types on the Revised Hubble Type
system was made by visual inspection of images displayed on a
workstation.  A purpose written script, running under IRAF and using
the XIMTOOL display, stepped through each entry in the catalog brighter
than the magnitude limit and displayed a pair of $100 \times 100$ pixel
grayscale images, stretched at 0--100DN and 0--400DN, using a
logarithmic display scaling.  We use logarithmic display to mimic the
response of the photographic plates, the source material used for the
fundamental schemes of galaxy classification.  By panning through the
lookup table, these display combinations offer a wide dynamic range
from outer wisps of low surface brightness to high-surface-brightness
core structure, allowing detailed and robust classification of the
object morphologies.  

Galaxy classifications include four components:  (1) Revised Hubble
type, (2) disturbance index -- the perceived disturbance of the galaxy
image, (3) dynamical state -- our interpretation of the cause of any
observed disturbance and (4) comments.  In addition to the standard
elliptical type, Morgan's ``D'' and ``cD'' types were assigned in some
obvious cases, though this has not been done in a rigorous way.
Revised Hubble types for spirals were assigned in half-class
increments, although it is questionable that there is sufficient
information in images sampled at this spatial resolution for such
discrimination.   Nevertheless, we retained such a system to
accommodate the scatter between the various classifiers.  Morphological
classifications of high redshift galaxies may differ from ground-based
observations of nearby systems because of several effects, including
(1) limited resolution and low signal-to-noise ratios -- particularly
for the smaller and fainter objects, (2) $(1+z)^4$ dimming resulting in
the loss of information in the fainter outer regions of galaxies, and
(3) differential K corrections between bulges and disks.  At present it
is unknown how these, or other biases, affect the comparison of
classifications of distant galaxies with better resolved, higher
signal-to-noise images of nearby galaxies, so caution must be exercised
in drawing conclusions that are based on more than rudimentary
differences.

S0 galaxies were seldom subtyped because of a similar lack of detail
(for example, small nuclear dust lanes).  Distinguishing elliptical
from face-on S0 galaxies is also challenging, particularly at the
signal-to-noise ratio and spatial sampling of these high-redshift
galaxy images (see Fig.~4). We have used the classes E/S0 or S0/E to
describe these ambiguities rather than to classify actual transition
cases (in this we differ from the approach used in the Revised Hubble
scheme); the order reflects whether the galaxy was more likely to be an
E or S0 in the classifier's opinion.  Disk galaxies with obvious bars
were so noted, for example, SBab, but again, because of limited spatial
sampling, their number is very incomplete and should not be taken as
representative of the true fraction of barred versus unbarred
galaxies.   Those objects not well described by an Revised Hubble type
were noted with ``?'' and those whose images appeared non-stellar but
were too small for reliable classification were denoted by ``X'' for
compact.

The visual inspections were used to make two additional assignments
other than the revised Hubble type.  It was felt that some indication
of the disturbance of the galaxy image would be useful and hence each
classifier recorded a disturbance index ($D$): 0 --- little or no
asymmetry, 1 or 2 --- moderate or strong asymmetry, 3 or 4 --- moderate or
strong distortion.  These were intended to be objective judgements
independent of the possible reason for the disturbance.  Accompanying
this was a subjective, interpretive description of the possible cause
of the disturbance.  For those objects where it seemed warranted the
following classes were assigned:  I --- tidal interaction with a
neighbor, M --- tidal interaction suggesting a merger, T --- tidal
feature without obvious cause, or C --- chaotic.  To the extent that
these are subjective and interpretive judgements, they should be viewed
with caution. For example, the ``merger'' category was assigned
whenever the structure was noticeably disturbed and one or more
apparently tidal tails were present, or in the case of two or more
close nuclei in a common envelope, but rarely were these as clear cut
examples of mergers such as N4038/9 (the so-called ``antenae'') where two
bodies and two tidal tails are clearly visible.  

All clusters were classified by AD and all but Cl1447$+$23 were
classified by WJC.  RSE classified Cl0016$+$16, Cl0054$-$27 and
Cl0412$-$65, and AO classified Cl0939$+$39, Cl0024$+$16 and
Cl1447$+$23.  Merging of the independent classifications was performed
by AD using a second IRAF script to redisplay each galaxy and the
alternative classifications, allowing him to merge the  classification
lists.  In $\geq 80$\% of cases to our classification limit the
classifiers agreed to one class or better, that is, the difference
between an E or S0, or an Sb and Sc (Fig.~1).  In these cases the modal
class was adopted, or, when only two classifications were available, AD
would alternate between the classifiers.  In those cases where
disagreement was more than one class, AD attempted to identify the
cause for the difference and resolve the classification in a consistent
manner.  In most cases there was sufficient information from the
individual classifiers for AD to make an objective resolution, but in a
small number of cases the final choice was his alone and hence
subjective.  Nevertheless, we believe that the uncertainty in this
process is small compared to the intrinsic ambiguity in assigning types
with the available images.   The comments included in Tables 4a-k not
only describe important features that went into the classification, but
also record the uncertain features that resulted in significant
disagreements between classifiers, for example ``faint protrusions top
and bottom, or disk?'' or ``spiral arms barely visible?''  In summary,
the majority of the morphological classifications are reliable to
better than one class, but in a few, predominantly fainter cases, where
features such as disks and arms are at the limit of detectability, the
uncertainty may be more than one class.

\vskip0.5truein
\centerline{\hbox{
\psfig{figure=f1.ps,height=4.0in,width=4.0in}
}}
{\footnotesize \addtolength{\baselineskip}{-5pt} 

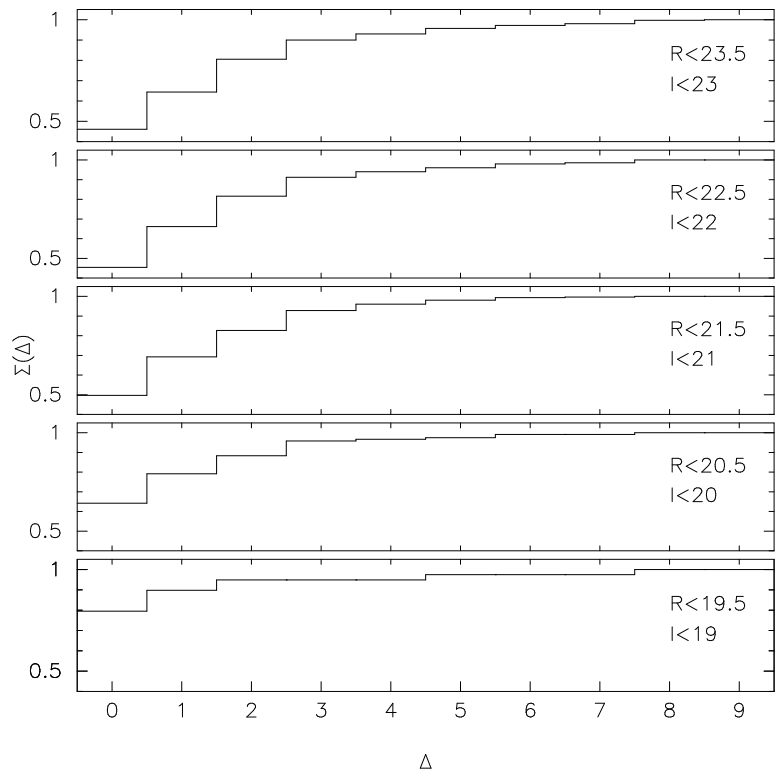
{\bf Figure 1.} The cumulative distribution of the absolute
differences in T types assigned to galaxies by independent pairs of
classifiers.  These are shown for a number of magnitude limits.  The
vertical scale of each plot runs from 40\% to 100\% in 10\% steps.  One
class in the Revised Hubble scheme is roughly equivalent to 2-3 units
on the T type scale.  We achieve agreement within one Hubble class for
more than 80\% of galaxies down to our completeness limit.

\addtolength{\baselineskip}{5pt}
}

The classifications from RSE are particularly important because of his
central role in the typing of galaxies in the Medium Deep Survey (MDS;
Griffiths et al.\ 1994), an intermediate-redshift ``field'' sample with
which we wish to compare.  For this reason we have been very critical
in our comparison of differences between RSE's classification and the
other 3 classifiers.  We judged that, while the scatter between RSE and
AD/WJC/AO is marginally larger than internal scatter of the latter
group, there are no systematic differences between the RSE (MDS) system
and the one adopted here.  At the level of the published MDS
classifications, which, for example, does not discriminate between E
and S0 galaxies and the full range of subclasses of spirals, we believe
that the data presented here can be considered to be on a consistent
system with the MDS.

The estimates of the disturbance index and the dynamical state were
merged by AD in a similar manner to the morphologies.  The disturbance
indices were found to agree to better than one unit, on average.  The
dynamical state is not continuous, nor has it been required
that each classifier supply one. Indeed, we find that more often
than not, in comparing pairs of observers, only $\sim$40\%
of cases are assigned a dynamical state by both.  Of these, however,
the classifiers agree on the descriptive index more than 60\% of the
time, far better than the chance occurence of 25\%.  The fact that
most objects were not flagged for comment is a clear indication
of the subjectivity of this index, but, at the same time, the 
agreement of the assigned category suggests that something genuine 
is being quantified in cases of obvious disturbance.

The final assignments of revised Hubble types were converted to de
Vaucouleurs ``T types'' as coded in Table 2b of the Second Reference
Catalog of Bright Galaxies (de Vaucouleurs, de Vaucouleurs, and Corwin
1976).   The main Hubble types are:  D/cD, $-$7; E, $-$5; S0,
$-$2; Sa, 1; Sb, 3; Sc, 5; Sd, 7; Sm, 9 and Irr, 10 (note $-$1 is not
used).  The full morphological information on our cluster fields is
given in Tables~4a-k.

\vskip0.2truein
\centerline{\hbox{
\psfig{figure=f2.ps,height=4.0in,width=4.0in}
}}
{\footnotesize \addtolength{\baselineskip}{-5pt} 

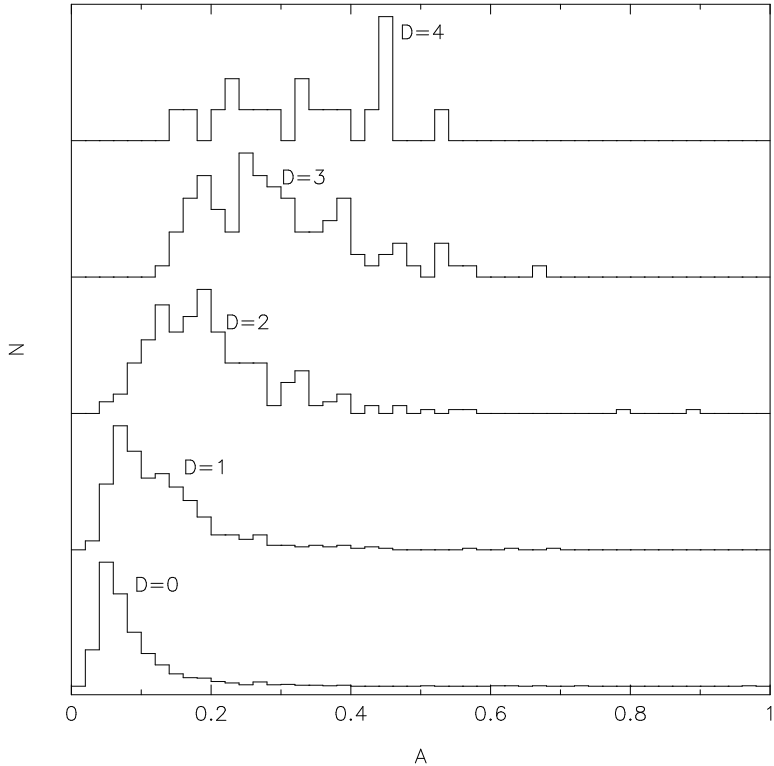
{\bf Figure 2.} The distribution of the algorithmically
determined asymmetry parameter ($A$) at a given value of the
disturbance index ($D$).  There is obviously a good correlation between
the two parameters. The scatter at a fixed  disturbance index is
similar in magnitude to the difference between successive disturbance
bins, supporting the resolution of the adopted disturbance scale.

\addtolength{\baselineskip}{5pt}
}

In addition to the visual estimates of morphological disturbance
determined from these samples, we have also measured a simple
quantitative index of image asymmetry.  This index measures the
proportion of light in a galaxy  in  an asymmetric component compared
to the total luminosity.  Operationally we determine this value by
rotating the galaxy by 180 degrees around its intensity-weighted
centroid and subtracting this rotated image from the original.  The
subtracted image is then smoothed with a 0.2 arcsec diameter top-hat
filter and the positive flux within the original detection isophote
summed, the ratio of this to the total isophotal flux of the
galaxy gives the asymmetry parameter ($A$).  

\subsection{Comparison of Visual and Machine-based Parameters}
 
In this section we compare various visually determined
properties of galaxy morphology with the equivalent algorithm-derived
parameter to highlight the strengths and weaknesses of the two
approaches.  

\vskip0.3truein
\centerline{\hbox{
\psfig{figure=f3.ps,height=3.0in,width=3.0in}
}}
{\footnotesize \addtolength{\baselineskip}{-5pt} 
%\vskip-0.2truein

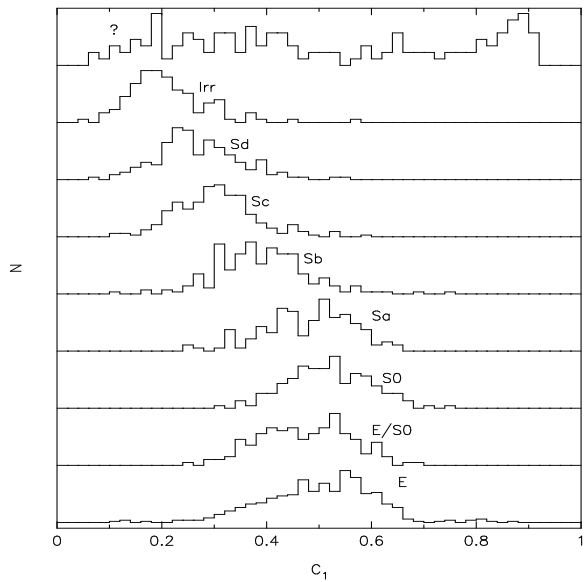
{\bf Figure 3.} The distribution of the concentration
index ($C_1$) for different morphological types showing the
good correlation between bulge-to-disk ratio (crudely measured
by $C_1$) and type for spiral galaxies.   For morphological types
of Sa and earlier, however, the concentration index fails to 
provide any useful differentiation.  The histogram at the
top of the plot shows the distribution of $C_1$ for unclassified
and compact objects on the frames.

\addtolength{\baselineskip}{5pt}
}

We start by showing in Fig.~2 the distribution of the asymmetry
parameter ($A$) for different values of the visually-determined
disturbance index ($D$).  There appears to be a good correlation
between these two, and furthermore the scatter in the asymmetry for a
given value of the disturbance index supports the resolution adopted in
our scheme.  The only instance where the disturbance index obviously
out-performs the quantitative asymmetry is when there are close
companions to the galaxy (lying within the detection isophote of the
object) or when the galaxy lies on a background with a strong
gradient.  Both of these are particularly a problem for the case of the
``additional'' objects (ID $> 3000$) which have been deblended by hand
and for this reason we suggest the asymmetry index for such
object should be treated with  caution.

Figure~3 shows the distribution of the concentration index
($C_1$) as a function of Hubble type for the full morphological
sample.  This comparison, as well as that between the algorithm-based
asymmetry parameter and the visual disturbance index shown in Fig.~2,
illustrates both the strengths and weaknesses of algorithm-based
classifications.   In particular, the concentration index, as used
here, becomes a useless discriminator for types Sa and earlier, which
have virtually identical distributions  of this quantity over a
substantial range in morphology.  Additional information from other simple
algorithm-based parameters (e.g.\ asymmetry) would obviously not remove
this failure.  This highlights the importance of more sophisticated
morphological parameterization, including visual inspection, in
understanding the evolution of galaxy structure and stellar
populations.

\hbox{~}\vskip0.3truein
\centerline{\hbox{
\psfig{figure=f4.ps,height=4.0in,width=3.0in}
}}
{\footnotesize \addtolength{\baselineskip}{-5pt} 
%\vskip-0.4truein
 
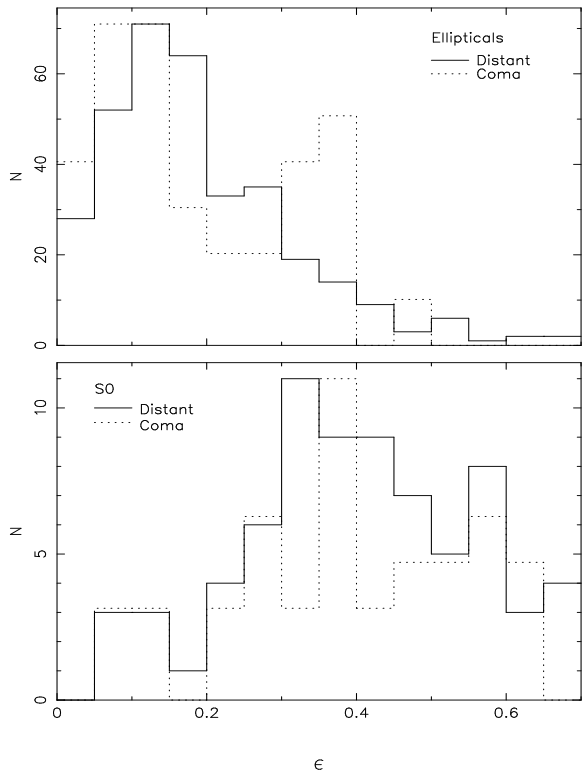
{\bf Figure 4.} The ellipticity distributions for E and S0 galaxies in
the distant clusters (solid histogram) compared to equivalent samples
drawn from the Coma cluster from Andreon et al.\ (1996; dotted
histogram).  We find good agreement between the shapes of the
distributions within each morphological type.  A more extensive
comparison with the equivalent distributions from the Revised
Shapley-Ames catalog (Sandage, Freeman \& Stokes 1970) and the cluster
catalog of Dressler (1980), both based on high-quality plate material,
indicate that at worst we are missing $\ls 50$\% of the face-on S0
galaxies (those with $\epsilon \leq 0.3$) amounting to $\sim 12$\% of
the total population.  The ordinate gives the number of galaxies in
each bin in the distant cluster sample, the samples from Coma have been
rescaled to the same peak.

\addtolength{\baselineskip}{5pt}
}

The discrimination between E and face-on S0 galaxies has been a
tradition concern in morphological classification.  While it is
desirable that this distinction exactly follow the definition of the
Revised Hubble system, which for face-on S0 galaxies means the ability
to discern a more extended outer envelope than would be seen for an
elliptical galaxy, it is equally important for our purposes that our
classifications be consistent with similar studies, particularly those
of low-redshift clusters imaged on photographic plates or with CCDs.
Accordingly, we have attempted to estimate the degree of incompleteness
in our S0 classification due to our missing face-on systems by
comparing the ellipticity distributions of our distant E and S0
galaxies with similar samples from both local rich clusters and the
field.  Perhaps the best sample for this purpose is the CCD survey of
Coma galaxies by Andreon et al.\ (1996) who provide axial ratios for a
magnitude-limited sample of $\sim 100$ galaxies in the central regions
of Coma.  These are CCD images, like those used here, measured at the
$\mu_R = 24.0$ mags arcsec$^{-2}$ isophote.  With $(1+z)^4$ dimming and
K corrections applied this isophote is equivalent to that used in our
analysis of the distant clusters, and due to the $\sim$ 2 mag darker
sky as seen from space, the contrast of this isophote with the sky is
nearly the same for the Andreon sample as our own HST sample.  We show
the distributions for the galaxies in Coma and our distant clusters in
Fig.~4.  A two sample Kolmogorov-Smirnov gives a 42\% probability that
the distant ellipticals and those in Coma have the same ellipticity
distribution, while the S0 samples agree at an even higher level
(88\%).  Hence at least from this comparison there is no strong
evidence of an enhanced deficit of round face-on S0's or a
corresponding increase in the proportion of round ellipticals in our
catalogs compared to similar samples of local clusters.

In terms of the absolute numbers of face-on S0's, the comparison given
above is misleading, in that Andreon et al.\ (1996) are probably also
missing a fraction of the face-on S0's in their sample.  This
conclusion comes from taking the simplest model of S0's as
randomly-projected thin disks, which would predict a flat distribution
of numbers of S0's with ellipticity.  To further investigate this we
have chosen to compare the fraction of round (arbitrarily chosen to be
$\epsilon \leq 0.3$) S0's in our catalog with the equivalent values
from the Revised Shapley-Ames catalog (Sandage, Freeman \& Stokes 1970;
SFS) and the cluster catalog of Dressler (1980), both based on very
high-quality plate material.  We find $23\pm6$\% of our S0's have
$\epsilon \leq 0.3$, compared to $33\pm4$\% for SFS and $34\pm3$\% for
Dressler (1980).  While not statistically significant, our value is
lower than these two measurements and would indicate that we could be
missing up to 40--50\% of the face-on S0's, or 12\% of the total S0
population.  This would seriously compromise the use of our catalog for
determining the intrinsic shapes of the S0 population in our distant
clusters (c.f.\ SFS), but does not substantially change the proportion
of S0's in the clusters.  In summary, it seems likely that our sample
{\it is} deficient in face-on S0 galaxies compared with a pure
application of the Revised Hubble system, but that this deficiency is
shared by most of the work involving the morphological classification
of all but the nearest galaxy clusters.

%\vfil\eject

\section{Cluster Galaxy Populations}

We now present some basic properties of the galaxy populations in the
clusters discussed here. We primarily discuss the morphological
composition of the clusters amd then touch briefly on a number of areas
which are dealt with in more depth in other publications.

We first discuss the morphological mixes in the distant clusters to a
fixed absolute magnitude, $M_V=-17.5 + 5 \log h$. Before doing this,
however, we have to correct the samples for field contamination.  Here
we rely on the extensive morphological survey of faint field galaxies
by the MDS team.  Using the catalogs kindly provided to us by the MDS
team we have fitted power laws to the differential number counts as a
function of morphological type  to $I_{814}=22.0$.  For the E and S0
subclasses (which are combined in the published MDS results) we use the
classifications from RSE which include these classes.  These are then
used in each of the cluster fields (extrapolating slightly when
necessary) to estimate the contamination from field galaxies (and its
morphological mix) in our catalogs. At $I_{814}=23.0$ we estimate the
field contamination is $\sim 11.2$ galaxies/arcmin$^2$, equivalent to
60 galaxies per WFPC-2 field.  The rough breakdown into morphological
classes is: E, 10\%; S0, 10\%; Sab, 25\%; Scdm, 30\% and Irr, 25\%.
These field distributions are then subtracted from the cluster
morphological distributions (assuming flat distributions in T type
within each of the larger MDS classification bins).  We show the
field-corrected distributions of galaxy morphology brighter than
$M_V=-17.5 + 5 \log h$ in our clusters in Fig.~5.  The absolute
magnitudes of the galaxies are determined from their ``Kron''
magnitudes (variable-diameter aperture magnitudes whose size depends
upon the galaxy scale size, see Bertin \& Arnouts (1996) and also Kron
(1980)), assuming that they lie at the cluster redshift.  We then use
the relevant non-evolving spectral energy distribution for the observed
morphological type to estimate the K correction.  This ignores any
possible luminosity dependence of the K correction within a
morphological class.  This effect should be negligible as we are
estimating luminosities from passbands which sample close to restframe
$V$.   Taking the ellipticals as an example we expect the variation in
K correction as a function of galaxy luminosity will lead to an error
in the estimated luminosity of less than 1\% across the range covered
by our samples.

The most immediate features of the distributions shown in Fig.~5 are
the dominant population of elliptical galaxies (T type $-$5) seen in
the central fields and the decline in the prevelance of this population
in the outskirts of the clusters (as indicated by the distributions for
the two outer fields imaged in our survey; A370\#2 and
Cl0939$+$47\#2).  It is obvious that even at a redshift of $z\sim 0.5$
the dominant bright galaxy component in the cores of rich clusters are
morphologically identifiable as elliptical galaxies.  Furthermore, we
also find that within the core fields the ellipticals are concentrated
around the cluster centers, with an average radial surface density
profile fitted by $\Sigma(r) \propto r^\gamma$ with
$\gamma=-0.80\pm0.10$ out to 200 h$^{-1}$ kpc, close to an isothermal
profile.  The profile is calculated around the cluster centers as
indicated by the lensing shear field (\cite{irs96}) and has been
corrected for uniform field contamination.   The evolution of the
``morphology-density'' and ``morphology-radius'' relations are
discussed in detail in Dressler et al.\ (1996).

\vfil\eject
\hbox{~}\vskip0.3truein
\centerline{\hbox{
\psfig{figure=f5.ps,height=7.0in,width=5.0in}
}}
{\footnotesize \addtolength{\baselineskip}{-5pt} 
%\vskip-0.4truein
 
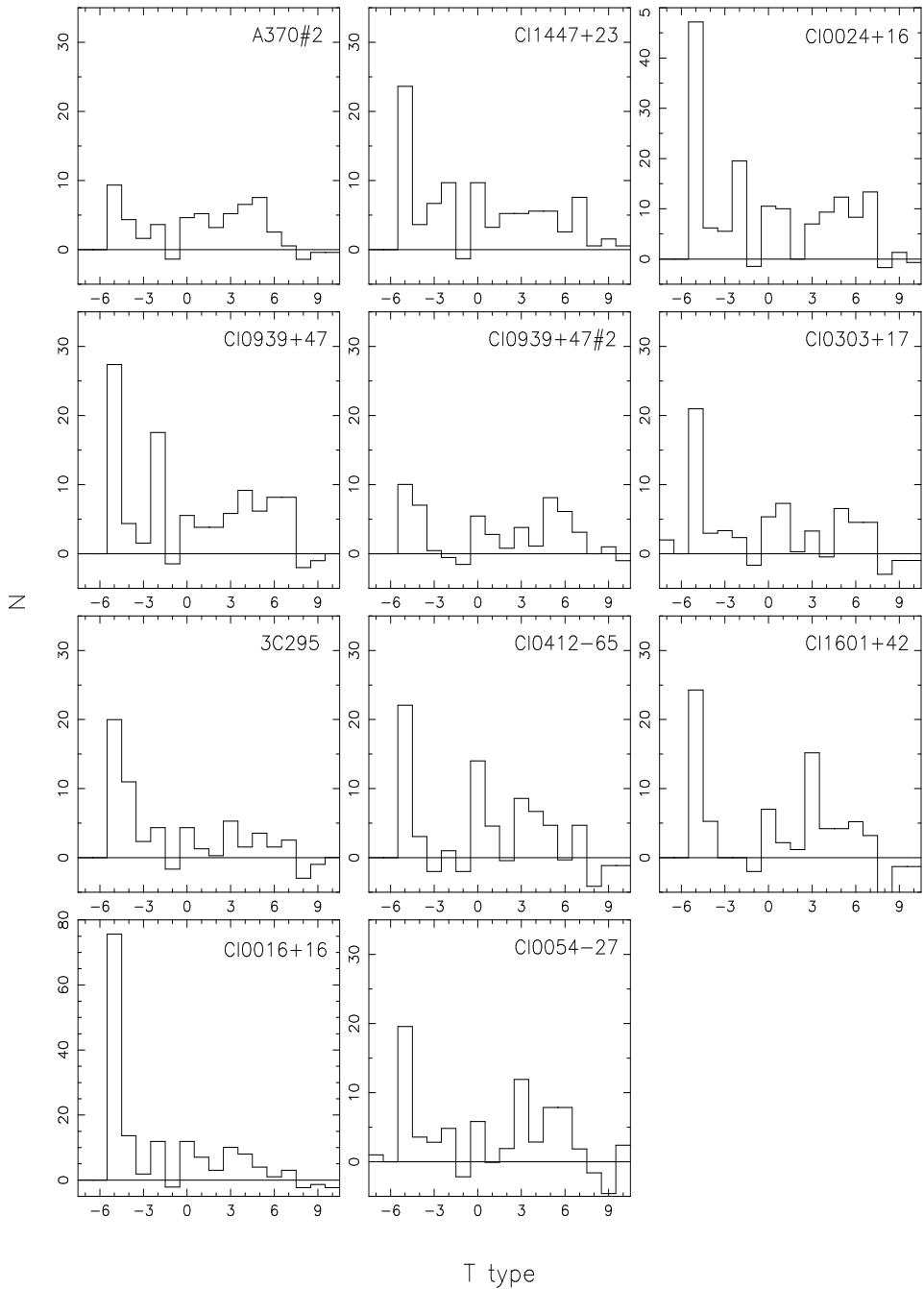
{\bf Figure 5.} The distribution of galaxy morphology
expressed as T types, brighter than $M_V=-17.5 + 5 \log h$ in our 11
fields, 9 central fields and the 2 outer fields A370\#2 and
Cl0939$+$47\#2.  These have been corrected for reddening and then field
contamination using the morphologically classified field counts from
the MDS.  Note the different vertical scales on some plots.

\addtolength{\baselineskip}{5pt}
}

We plot the luminosity function of the elliptical populations in our
clusters in Fig.~6.  The galaxy luminosities and field corrections have
been estimated in the same manner as was used for the morphological
mixes.  The luminosity functions have then been combined into 3
independent redshift bins: A370\#2, Cl1447$+$23 and Cl0024$+$16 with
$<\!  z\! >=0.38$; Cl0939$+$47, Cl0939$+$47\#2, Cl0303$+$17 and 3C295
with $<\!  z\! >=0.43$; Cl0412$-$65, Cl1601$+$42, Cl0016$+$16 and
Cl0054$-$27 with $<\!  z\! >=0.54$.  Fitting a standard Schechter
function to these distributions, with a fixed faint end slope of
$\alpha=-1.25$, we obtain characteristic luminosities of $M^\ast_V =
-20.47\pm0.18 + 5 \log h$ at $<\! z\! >=0.38$, $M^\ast_V =
-20.62\pm0.13 + 5 \log h$ ($<\!  z\! >=0.43$) and $M^\ast_V =
-20.76\pm0.12 + 5 \log h$ for $<\! z\! >=0.54$.  These fits are shown
overlaid on Fig.~6 and the quoted limits include the uncertainty in the
field contribution to the counts.  For a low redshift comparison we
take the analysis by Colless (1989) of 14 rich clusters, where he
derives $M^\ast_{b_j} = -19.84\pm0.06 + 5 \log h$ for $\alpha=-1.25$,
or $M^\ast_V \sim -20.5 + 5 \log h$ assuming a mean color of
$(b_j-V)\sim0.7$.  Thus, there appears to be a slow, but significant
brightening of the population between $z=0.0$--0.54 amounting to at
least $\delta M_V \sim -0.3$.  This brightening is consistent with
observations of the homogeneity of the colors of these galaxies (Ellis
et al.\ 1996) and their luminosity-size relationship (Barger et
al.\ 1996), both of which indicate that the bulk of the stars in this
population were formed at very early epochs.  

Figure~6 shows that although we have not allowed the faint end slope of
the elliptical population to vary it does appear to be reasonably
described by the local value of $\alpha\sim -1.25$.  We highlight this
because of recent discussions of the relative importance of age and
metallicity effects in defining the color-magnitude relation of cluster
elliptical galaxies (Worthey 1994; Kodama \& Arimoto 1996).  If age
differences are assumed to cause the trend of fainter ellipticals being
bluer then Kodama \& Arimoto (1996) found that to adequately describe
the local color-magnitude relation $L^\ast$ ellipticals are still
forming at $z\sim 0.2$--0.3 ($h=0.5$ and $q_o=0.5$).  They point out
that the color evolution implied by this model is incompatible with the
observed form of the color-magnitude relation in distant clusters.
Equally, in such a model the bulk of the elliptical population fainter
than $\sim L^\ast$ would not be present in the most distant clusters we
observe.   We would thus expect the ratio of the numbers of low and
high luminosity ellipticals to be considerably lower than in local
clusters, corresponding to a shallower faint end slope.  The absence of
any strong change in $\alpha$ for ellipticals brighter than $L_V\sim
L^\ast + 3$ out to $z\sim 0.6$ would appear to contradict this. We thus
confirm Kodama \& Arimoto's conclusion that age variations are unlikely
to be the dominant cause of the color-magnitude relation of ellipticals
in rich clusters.

Returning to Fig.~5 we also note that the clusters contain only small
populations of S0 galaxies ($\sim 15$\%).  We have shown in Fig.~4 that
this is unlikely to have resulted from a widespread misclassification
of S0's as elliptical galaxies, although our misclassification of
face-on S0's will lead to an increase in the total S0 population by
$\sim 12$\% -- raising the fraction of S0 galaxies to $\sim 17$\% of
the total cluster population.  The absence of the S0 population is thus
particularly intriguing, given that they constitute $\sim 40$\% of
local clusters.  Unfortunately, with so few S0's it is impossible to
obtain adequate numbers to measure the luminosity function and its
evolution within our sample.   The small fractions of S0's in our
clusters and the implications of this for their formation and evolution
is outside the scope of the discussion here and we return to it
elsewhere (Dressler et al.\ 1996).  We note that the S0's lie
intermediate between the ellipticals and the spirals (see below) in
their radial distribution, with $\gamma=-0.59\pm0.15$ interior to 200
h$^{-1}$ kpc.  

Moving to later Hubble types we see that all clusters contain
populations of these systems.  The field-corrected spiral fraction in
the clusters (the fraction of the total population which have
morphologies of Sa or later) is listed in Table~1. The radial profile
of these galaxies around the cluster centers is substantially shallower
than the elliptical population, dropping only as $\gamma=-0.29\pm0.08$,
as expected for a dynamically unrelaxed component of the clusters.
Following our earlier discussion we plot the combined luminosity
function of the cluster spirals (Sa's and later, identified
statistically by subtracting the field counts from the MDS) in Fig.~7.
Again fitting a Schechter function, this time with $\alpha=-1.0$, we
see no evidence for change between the different epochs, in contrast to
the elliptical  populations.  We find $M^\ast_V = -20.19\pm0.18 + 5
\log h$ at $<\! z\! >=0.38$, $M^\ast_V = -20.17\pm0.15 + 5 \log h$
($<\!  z\! >=0.43$) and $M^\ast_V = -20.21\pm0.15 + 5 \log h$ for $<\!
z\! >=0.54$.   We compare these with the parametrisation of the field
galaxy luminosity function by Lilly et al.\ (1995), which is dominated
by spiral galaxies.  They find $\alpha=-1.0$ and $M^\ast_B\sim -19.4 +
5 \log h$ for galaxies brighter than $M^\ast_B<-17.2 + 5 \log h$ in the
redshift range $z=0.2$--0.5, roughly equivalent to that probed here.
Taking a mean restframe color of $(B-V)\sim 0.6$ for this population we
have $M^\ast_V\sim -20.0 + 5 \log h$, very close to that found in our
sample.  Given the differences in the photometric systems used in the
cluster and field surveys, as well as the different sample selection
criteria, we feel that it is too early yet to claim a systematic
difference between the luminosities of spiral galaxies in distant
clusters and the surrounding field. 

Although the HST images discussed here provide excellent photometry,
with small photometric errors for even the fainter galaxies (thanks to
the significantly darker sky background from space), uncertainties in
the luminosity functions derived for the clusters remain large because
of the substantial contamination by foreground and background galaxies
--- a much more significant problem than for nearby clusters.  The
advent of the new generation of large, ground-based telescopes will
considerably improve the situation by providing extensive membership
information for galaxies in these fields.  With large samples of
confirmed cluster members it will be possible to robustly measure the
evolutionary brightening of the elliptical population, while comparison
of the luminosity function of spiral members with that derived for the
field will indicate whether there are any substantial differences
between these two populations.   Such an effort, though intensive of
telescope time, would be very worthwhile. 

\vfil\eject

\hbox{~}\vskip0.7truein
\centerline{\hbox{
\psfig{figure=f6.ps,height=3.0in,width=5.6in}
}}
{\footnotesize \addtolength{\baselineskip}{-5pt} 

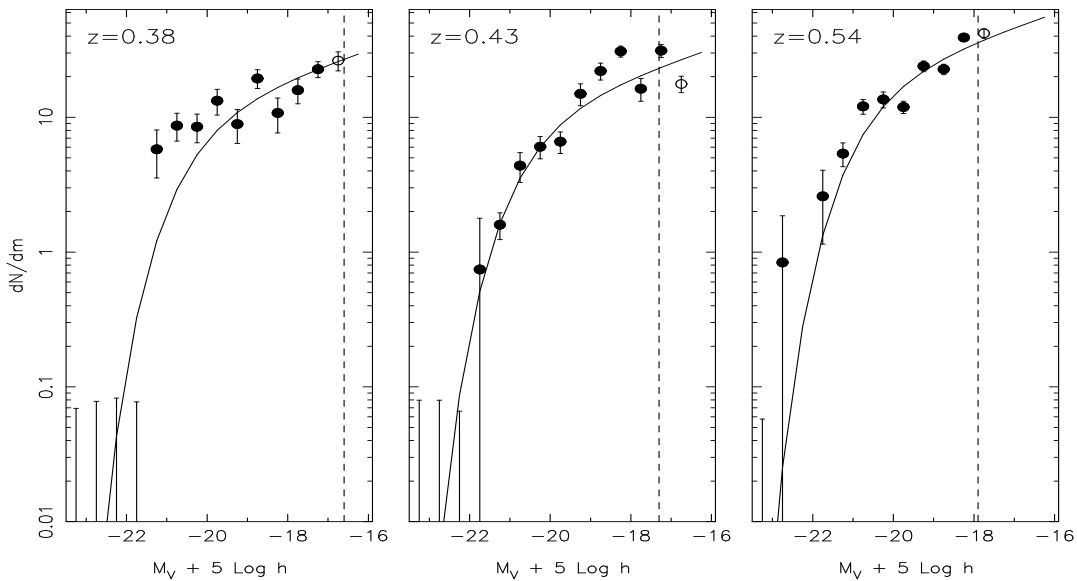
{\bf Figure 6.} The luminosity function of cluster elliptical
galaxies in three redshift bins. The counts are corrected
for field contamination and then converted to absolute $V$ magnitudes
assuming K corrections for a non-evolved elliptical spectral
energy distribution.  Overlayed on these are the best fitting
Schechter function (with a fixed $\alpha=-1.25$ faint end slope).
Points with filled symbols are used in the fitting.  The dashed line
shows the magnitude limit for the samples.  

\addtolength{\baselineskip}{5pt}
}

\hbox{~}\vskip0.3truein
\centerline{\hbox{
\psfig{figure=f7.ps,height=3.0in,width=5.6in}
}}
{\footnotesize \addtolength{\baselineskip}{-5pt} 

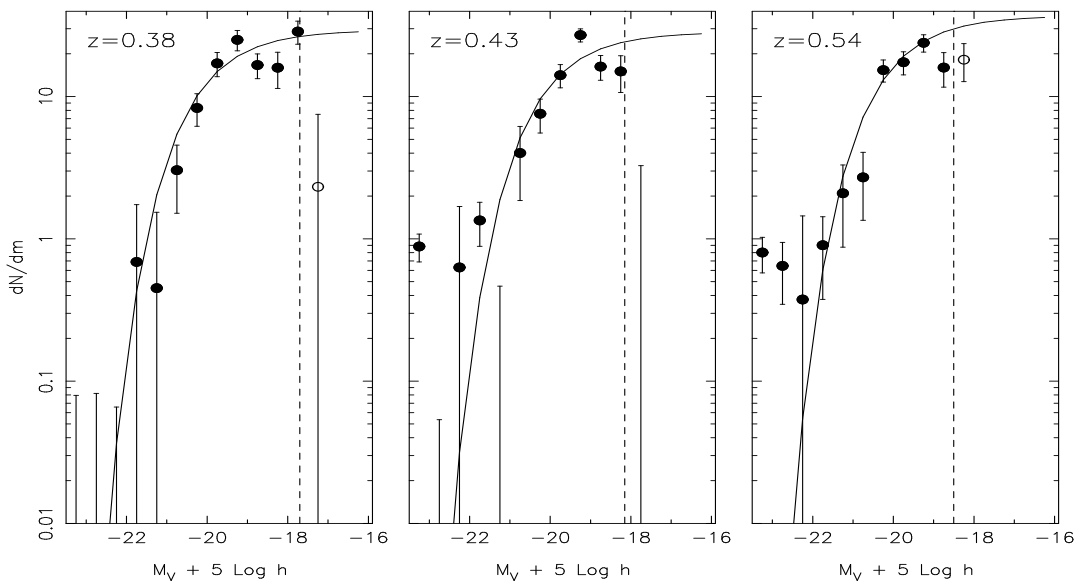
{\bf Figure 7.} The luminosity function of cluster spiral galaxies in
three redshift bins (the same as those used in Fig.~6). The counts are
corrected for field contamination and then converted to absolute $V$
magnitudes assuming K corrections for the relevant non-evolved spectral
energy distribution of each morphological sub-type.
Overlayed on these are the best fitting Schechter function (with a
fixed $\alpha=-1.0$ faint end slope).  Points with filled symbols are
used in the fitting.  The dashed line shows the magnitude limit for the
samples.

\addtolength{\baselineskip}{5pt}
}

\vfil\eject

\section*{Acknowledgements}
We thank Ray Lucas at STScI for his enthusiastic help which enabled the
efficient gathering of these HST observations and Richard Griffiths for
the use of the MDS catalog.  We also wish to thank the referee,
Dr.\ Allan Sandage, for his thorough reading and lucid comments on this
paper, especially in regard to our morphological classifications.
Finally, we thank Alfonso Arag\'on-Salamanca, Nobuo Arimoto, Amy
Barger, Bianca Poggianti and Gillian Wilson for useful discussions and
assistance.  IRS, RSE and RMS acknowledge support from the Particle
Physics and Astronomy Research Council. AD and AO acknowledge support
from NASA through STScI grant 3857.   WJC acknowledges support from the
Australian Department of Industry, Science and Technology, the
Australian Research Council and Sun Microsystems and also an ESO
Visiting Fellowship during part of this work.

%\vfil\eject

\vfil\eject
\doublespace

\centerline{\bf PLATE CAPTIONS}

\noindent{\bf Plate 1a}. Field of A370\#2.

\noindent{\bf Plate 1b}. Field of Cl1447$+$23.

\noindent{\bf Plate 1c}. Field of Cl0024$+$16.

\noindent{\bf Plate 1d}. Field of Cl0939$+$47.

\noindent{\bf Plate 1e}. Field of Cl0939$+$47\#2.

\noindent{\bf Plate 1f}. Field of Cl0303$+$17.

\noindent{\bf Plate 1g}. Field of 3C295.

\noindent{\bf Plate 1h}. Field of Cl0412$-$65.

\noindent{\bf Plate 1i}. Field of Cl1601$+$42.

\noindent{\bf Plate 1j}. Field of Cl0016$+$16.

\noindent{\bf Plate 1k}. Field of Cl0054$-$27.

\singlespace

\vfil\eject

\thispagestyle{empty}
\begin{sidetable}

\centerline{\sc \hfil Table 1 \hfil }

\centerline{\sc \hfil Cluster sample and properties \hfil }

{\small\scriptsize
\hspace{-1.truein}\begin{tabular}{lccrccccccccccccc}
\noalign{\medskip}
\hline\hline
\noalign{\smallskip}
{Cluster} & {R.A.} & {Dec.} & {P.A.} & {$z$} & \multispan{4}{\hfil T$_{\rm exp}$~(ks) \hfil}  & {L$_X${\scriptsize (0.3--3.5)}} &
$\sigma$ \hfil &   {kpc/arcsec} & {\scriptsize E(B-V)} & {$\mu (1\sigma)$} & \multispan2{\hfil N \hfil}  & $f_{sp}$ \cr
& {\scriptsize (J2000)} & {\scriptsize (J2000)} & {\scriptsize V3 (deg)} &  & {\scriptsize F450W} & {\scriptsize F555W} & {\scriptsize F702W} & {\scriptsize F814W} & {\scriptsize h$^{-2}$ 10$^{\scriptsize 44}$ ergs} & {\scriptsize km/s [N]}&    (h$^{-1}$) & & & Morph & All & \cr
\noalign{\smallskip}
\noalign{\hrule}
\noalign{\smallskip}
A370\#2 & 02~40~01.1 & $-$01~36~45 & 235.00 & 0.37 & ... & 8.0 & ... & 12.6 & 2.73 &  1350 [34]&  3.09  & 0.01 & 27.5 & 129 & 542 & 0.54 \cr
Cl1447+23 & 14~49~28.2 & $+$26~07~57 & 125.35 & 0.37 & ...  & ... & 4.2 & ... & ... &  ...\hfil &  3.09 & 0.02 & 27.5 & 210 & 792 & 0.47 \cr
Cl0024+16 & 00~26~35.6 & $+$17~09~43 & 328.30 & 0.39 & 23.4  & ... & ... & 13.2 & 0.55 & 1339 [33]&  3.17 & 0.03 & 27.5 & 248 & 835 & 0.40 \cr
Cl0939+47 & 09~43~02.6 & $+$46~58~57 & 47.32 & 0.41 & ... & ... & 21.0 & ... &  1.05 & 1081 [31]&  3.26 & 0.00 & 27.3 & 207 & 975 & 0.46 \cr
Cl0939+47\#2 & 09~43~02.5 & $+$46~56~07 & 315.00 & 0.41 & ... & 4.0 & ... & 6.3 & 1.05 & 1081 [31]&  3.26 & 0.00 & 26.9 & 116 & 627 & 0.62 \cr
Cl0303+17 & 03~06~15.9 & $+$17~19~17 & 50.13 & 0.42 & ... & ... & 12.6 & ... & 1.05 & 1079 [21]&  3.29  & 0.12 & 27.9 & 161 & 891 & 0.46 \cr
3C295 & 14~11~19.5 & $+$52~12~21 & 282.04 & 0.46 & ... & ... & 12.6 & ... & 3.20 & 1670 [21]&  3.43  & 0.00 & 27.1 & 142 & 899 & 0.28 \cr
Cl0412$-$65 & 04~12~51.7 & $-$65~50~17 & 174.00 & 0.51 & ... & 12.6 & ... & 14.7 & 0.08 & ...\hfil &  3.59 & 0.00 & 27.5 & 155 & 865 & 0.47 \cr
Cl1601+42 & 16~03~10.6 & $+$42~45~35 & 315.00 & 0.54 & ...  & ... & 16.8 & ... & 0.35 & 1166 [27]&  3.67  & 0.00 & 28.3 & 145 & 928 & 0.46 \cr
Cl0016+16 & 00~18~33.6 & $+$16~25~46 & 250.00 & 0.55 & ... & 12.6 & ... & 16.8 & 5.88 & 1703 [30]&  3.67  & 0.03 & 26.7 & 211 & 837 & 0.21 \cr
Cl0054$-$27 & 00~56~54.6 & $-$27~40~31 & 121.00 & 0.56 & ... & 12.6 & ... & 16.8 &  0.25 & ...\hfil &  3.72 & 0.00 & 27.4 & 133 & 767 & 0.42 \cr
\end{tabular}
}
\end{sidetable}

\cleardoublepage

%\begin{center}
\centerline{\sc \hfil Table 2a \hfil }

\centerline{\sc \hfil Notes on parameters in Table 3 \hfil }

{\scriptsize
\hspace{-1truein}\begin{tabular}{rlllll}
\noalign{\medskip}
\hline\hline
\noalign{\smallskip}
{Column} & {Heading}  & {Parameter} & {Units} & {Format} & {Comment} \cr
\noalign{\smallskip}
\noalign{\hrule}
\noalign{\smallskip}
{1} & {ID} & NUMBER &  & i4 & Object identifier$^1$   \cr
{2} & {X} & X\_IMAGE & pixels & f8.2 & X centroid on frame  \cr
{3} & {Y} & Y\_IMAGE & pixels & f8.2 & Y centroid on frame   \cr
{4} & $I^{ap}_{814}$ & F814W\_APER & mags & f7.3 &  1.0$''$ diameter aperture magnitude$^2$   \cr
{5}  & $\delta I^{ap}_{814}$ & F814W\_ERR\_APER & mags & f7.3 & Error on above   \cr
{6} & $I^{iso}_{814}$  & F814W\_ISO & mags & f7.3 &  Isophotal magnitude above detection isophote  \cr
{7} & $\delta I^{iso}_{814}$ & F814W\_ERR\_ISO & mags & f7.3 &  Error on above  \cr
{8} & $I^{best}_{814}$ & F814W\_BEST & mags & f7.3 &  Optimal measure of total magnitude$^3$  \cr
{9}  & $\delta I^{best}_{814}$ & F814W\_ERR\_BEST & mags & f7.3 &  Error on above  \cr
{10} & {Area} & ISOAREA\_IMAGE & pixels & i5 &  Isophotal area above detection isophote  \cr
{11} & $r_{k}$ & KRON\_RADIUS & pixels & f6.2 &  Kron radius  \cr
{12} & $a$ & A\_IMAGE & pixels & f6.2 &  2nd order moment along the major axis  \cr
{13} & $b$ & B\_IMAGE & pixels & f6.2 &  2nd order moment along the minor axis  \cr
{14} & $\theta$ & THETA\_IMAGE & degrees & f6.1 &  Orientation of major axis$^4$   \cr
{15} & $\mu_{thresh}$ & MU\_THRESHOLD & mags arcsec$^{-2}$ & f7.3 &  Detection threshold \cr
{16} & $\mu_{max}$ & MU\_MAX & mags arcsec$^{-2}$ & f7.3 &  Peak surface brightness  \cr
{17} & {Sky} & BACKGROUND & DN & f8.2 &  Sky background level  \cr
{18} & {FWHM} & FWHM\_IMAGE & pixels & f6.2 &  FWHM from gaussian fit  \cr
{19} & $P_{star}$ & CLASS\_STAR &  & f5.2 &  Star-galaxy classification$^5$  \cr
{20} & {C$_1$} & CONC &  & f6.3 &  Concentration index$^6$  \cr
{21} & {C$_2$} & CONTRAST &  & f5.2 &  Contrast index$^6$  \cr
{22} & {Err} & FLAGS &  & i4 &  Error flags$^7$  \cr
{23} & $V_{555}-I_{814}$ & F555W$-$F814W & mags & f7.3 &  Color measured in a 1.0$''$ diameter aperture$^8$  \cr
{24} & $\delta(V_{555}-I_{814})$ & F555W$-$F814W\_ERR & mags & f7.3 &  Error on above$^8$  \cr
\end{tabular}
}
%
%\end{center}
%

\cleardoublepage

\begin{center}
{\sc \hfil Table 2b \hfil }

{\sc \hfil Notes on parameters in Table 4 \hfil }

{\scriptsize
\begin{tabular}{rllll}
\noalign{\medskip}
\hline\hline
\noalign{\smallskip}
{Column} & {Heading}  & {Parameter} &  {Format} & {Comment} \cr
\noalign{\smallskip}
\noalign{\hrule}
\noalign{\smallskip}
{1} & {ID } & NUMBER &   i4 &  Object identifier$^1$  \cr
{2} & {DG \# } & DG\_NUMBER &   a3 &  Cross identification to DG92$^9$  \cr
{3} & $I_{814}$ & F814W\_BEST &  f7.3 & Optimal measure of total magnitude$^{2,3}$    \cr
{4} & $A$ & ASYMMETRY &   f6.3 &  Asymmetry parameter$^{10}$  \cr
{5} & {Class} & MORPHOLOGY &   a12 & Hubble class$^{11}$   \cr
{6} & {T-type} & T\_TYPE &   i3 & T-type$^{12}$   \cr
{7} & {D} & DISTURBANCE &   i3 &   Disturbance index$^{13}$ \cr
{8} & {Int} & INTERPRETATION &   a8 &  Interpretation of disturbance$^{14}$  \cr
{9} & {Comments} & COMMENTS &   a62 &  Description of object morphology  \cr
\end{tabular}
}
\end{center}

\bigskip
{\small
{\scriptsize
\begin{tabular}{rrl}
\noalign{\medskip}
1 & \multispan{2}{ID's starting 2000 denote objects lying on the PC chip, \hfil}\cr
 & \multispan{2}{those starting from 3000 are ``additional'' objects lying on the WFC (see text).  \hfil}\cr 
2 & \multispan{2}{Parameter is F702W where applicable.  \hfil}\cr
3 & \multispan{2}{Variable-diameter aperture magnitude measured in an elliptical aperture of major axis radius $2.5 r_k$, unless  \hfil}\cr
  & \multispan{2}{Err=1 occured in which case it the ``corrected'' isophotal magnitude is used (\cite{eb95}).\hfil} \cr
4 & \multispan{2}{Defined counter-clockwise from the positive X axis in the range $-$90 to 90. \hfil}\cr
5 & \multispan{2}{Star-galaxy classifier using neural-net weights: 0.0 denotes a galaxy and 1.0 a star.  \hfil}\cr
6 & \multispan{2}{See text for details.  \hfil}\cr
7 & \multispan{2}{Error flag values and the associated error are quoted below:  \hfil} \cr
& 1   & --- bright neighbors may bias magnitude estimate, \cr
& 2   & --- originally a blend, \cr 
& 4   & --- saturated, \cr
& 8   & --- object truncated by frame boundary, \cr
& 16  &  --- incomplete aperture photometry, \cr
& 32  &  --- incomplete isophotal photometry, \cr
& 64  &  --- memory overflow during deblending, \cr
& 128 &   --- memory overflow during extraction. \cr
8 & \multispan{2}{Only for those fields with color information (F450W$-$F814W for Cl0024+16), \hfil} \cr
  & \multispan{2}{a value of 99.999 in these fields denotes an undefined measurement.  \hfil}\cr
9 & \multispan{2}{Non-exhaustive cross identification of objects against the catalogs in Dressler \& Gunn (1992),}\cr
  & \multispan{2}{for the clusters Cl0024+16, Cl0939+47, Cl0303+17, 3C295, Cl1601+42 and Cl0016+16.  \hfil} \cr
10 & \multispan{2}{An error is denoted by a value of 9.999, values for ``additional'' objects should be  \hfil}\cr
  & \multispan{2}{viewed with caution.\hfil}\cr
11 & \multispan{2}{The standard Hubble classification scheme (E, S0, Sa, Sb...) with the addition of:  \hfil}\cr
   & D, cD &  --- Morgan type D or cD galaxy, \cr
   & E/S0 or S0/E &  --- cannot distinguish E or S0,\cr 
   & X & --- compact object (likely non-stellar but too compact to see structure),\cr
   & $\ast$ &  --- stellar image, \cr
   & ?  & --- unclassifiable. \cr
12 & \multispan{2}{Standard T-type, an undefined entry is given as $-$99.  \hfil} \cr
13 & \multispan{2}{Disturbance index:  \hfil}\cr
& 0 & --- normal,\cr
& 1 & --- moderate asymmetry,\cr
& 2 & --- strong asymmetry,\cr
& 3 & --- moderate distortion,\cr
& 4 & --- strong distortion,\cr
& $-$99 & --- undefined.\cr
14 & \multispan{2}{Interpretation of disturbance classes:  \hfil}\cr
& M & --- merger,\cr
& I & --- tidal interaction with neighbor,\cr
& T & --- tidal feature,\cr
& C & --- chaotic,\cr
& ! & --- remarkable.\cr
\end{tabular}
}
}
\vfil\eject

\thispagestyle{empty}
\begin{sidetable}

\centerline{\sc \hfil Table 3A\footnote{Complete versions of
Table 3 (Tables 3A--3K) are available on AAS CD-ROM series, Vol. ZZZ}
 \hfil }

\centerline{\sc \hfil Sample Object Catalog for A370\#2 \hfil }
{\scriptsize
\hspace{-1.93truein}\begin{tabular}{rrrrrrrrrrrrrrrrrrrrrrrr}
\noalign{\medskip}
\hline\hline
\noalign{\smallskip}
{ID} & {X} &  {Y} & $I^{ap}_{814}$ & $\delta I^{ap}_{814}$ & $I^{iso}_{814}$ & $\delta I^{iso}_{814}$ 
& $I^{best}_{814}$ & $\delta I^{best}_{814}$ & {Area} & $r_{k}$ & $a$ & $b$ & $\theta$ & $\mu_{thresh}$ &
$\mu_{max}$ & {Sky} & {FWHM} & $P_{star}$ & {C$_1$} & {C$_2$} & {Err} & $V-I$ & 
$\delta(V-I)$ \cr
\hline
\noalign{\smallskip}
         1 &   1001.05 &      2.58 &    25.840 &     0.187 &    26.198 &     0.133 &    26.011 &     0.174 &         9 &      4.36 &      0.92 &      0.77 &      85.2 &    24.600 &    22.947 &     30.19 &      2.68 &      0.01 &     0.228 &      0.64 &        24 &    99.999 &    99.999  \cr 
         2 &    346.29 &      5.21 &    26.007 &     0.244 &    26.159 &     0.132 &    25.887 &     0.180 &        10 &      4.17 &      1.07 &      0.81 &      17.5 &    24.600 &    23.170 &     30.12 &      5.56 &      0.01 &     0.101 &      0.57 &        16 &    99.999 &    99.999  \cr 
         3 &   1040.08 &      4.20 &    25.306 &     0.128 &    25.515 &     0.103 &    25.085 &     0.144 &        17 &      5.00 &      1.84 &      1.34 &     -59.4 &    24.600 &    22.857 &     30.18 &      9.37 &      0.00 &     0.032 &      0.71 &        24 &    99.999 &    99.999  \cr 
         4 &   1483.05 &      5.95 &    22.511 &     0.017 &    22.182 &     0.016 &    22.123 &     0.017 &       140 &      4.17 &      4.11 &      1.67 &      27.3 &    24.600 &    20.989 &     30.78 &      9.01 &      0.02 &     0.283 &      0.74 &        24 &     1.347 &     0.041  \cr 
         5 &     81.31 &     11.73 &    25.497 &     0.157 &    26.278 &     0.149 &    25.346 &     0.175 &        11 &      6.27 &      1.34 &      0.80 &     -37.9 &    24.600 &    23.322 &     30.26 &      4.94 &      0.00 &    -0.002 &      0.64 &         0 &     2.351 &     0.699  \cr 
         6 &   1183.68 &      8.34 &    24.263 &     0.056 &    23.913 &     0.046 &    23.568 &     0.059 &        75 &      4.33 &      4.11 &      1.76 &     -37.7 &    24.600 &    22.972 &     30.22 &     16.09 &      0.00 &     0.135 &      0.52 &        24 &     1.018 &     0.106  \cr 
         7 &    438.81 &      4.99 &    20.123 &     0.005 &    20.080 &     0.005 &    20.121 &     0.005 &       167 &      3.50 &      1.59 &      1.31 &      12.2 &    24.600 &    16.524 &     30.19 &      1.43 &      0.95 &     0.818 &      0.95 &        24 &     2.684 &     0.018  \cr 
         8 &    629.91 &     10.39 &    24.349 &     0.060 &    24.544 &     0.060 &    24.107 &     0.072 &        37 &      4.83 &      1.85 &      1.52 &      62.0 &    24.600 &    22.437 &     30.33 &      5.23 &      0.00 &     0.207 &      0.64 &         0 &     1.840 &     0.220  \cr 
         9 &    311.50 &     17.81 &    25.628 &     0.176 &    25.951 &     0.121 &    25.592 &     0.130 &        14 &      3.69 &      1.49 &      0.67 &     -61.2 &    24.600 &    23.185 &     30.16 &      4.95 &      0.00 &     0.097 &      0.59 &         0 &     0.516 &     0.307  \cr 
        10 &    812.81 &     19.94 &    25.400 &     0.144 &    26.454 &     0.172 &    25.230 &     0.172 &         7 &      6.71 &      1.44 &      0.79 &      88.2 &    24.600 &    22.938 &     30.31 &      4.81 &      0.02 &     0.064 &      0.84 &         0 &     2.109 &     0.654  \cr 
        11 &   1506.88 &     19.29 &    25.412 &     0.146 &    25.694 &     0.109 &    25.351 &     0.150 &        14 &      5.10 &      1.56 &      0.74 &     -80.1 &    24.600 &    22.511 &     30.78 &      6.43 &      0.10 &     0.144 &      0.82 &         0 &     2.758 &     1.038  \cr 
        12 &   1284.62 &     23.99 &    25.277 &     0.130 &    25.744 &     0.109 &    25.073 &     0.130 &        15 &      6.00 &      1.46 &      0.72 &     -32.5 &    24.600 &    22.970 &     30.28 &      4.43 &      0.00 &     0.251 &      0.68 &         0 &     0.696 &     0.202  \cr 
        13 &   1310.76 &     24.57 &    25.603 &     0.172 &    26.289 &     0.151 &    25.617 &     0.162 &        10 &      4.64 &      1.19 &      0.84 &      85.8 &    24.600 &    23.392 &     30.25 &      4.30 &      0.00 &     0.173 &      0.68 &         0 &     0.705 &     0.261  \cr 
        14 &    998.20 &     26.30 &    25.481 &     0.154 &    26.393 &     0.159 &    25.337 &     0.151 &         9 &      6.04 &      0.94 &      0.91 &     -75.0 &    24.600 &    23.618 &     30.12 &      5.05 &      0.00 &     0.145 &      0.62 &         0 &     1.566 &     0.405  \cr 
        15 &     74.95 &     23.55 &    23.903 &     0.043 &    23.910 &     0.042 &    23.681 &     0.058 &        54 &      4.72 &      2.23 &      1.73 &     -41.7 &    24.600 &    21.734 &     30.25 &      5.56 &      0.02 &     0.240 &      0.68 &         1 &     1.129 &     0.089  \cr 
        16 &    276.89 &     17.36 &    23.089 &     0.024 &    22.488 &     0.021 &    22.330 &     0.027 &       191 &      4.43 &      5.90 &      2.07 &     -62.5 &    24.600 &    21.809 &     30.23 &     16.88 &      0.00 &     0.246 &      0.65 &        24 &     0.802 &     0.041  \cr 
        17 &    884.88 &     25.49 &    22.318 &     0.015 &    22.034 &     0.014 &    21.966 &     0.016 &       157 &      3.96 &      3.61 &      1.87 &     -44.3 &    24.600 &    20.710 &     30.18 &      6.68 &      0.02 &     0.344 &      0.71 &         0 &     0.911 &     0.027  \cr 
        18 &   1408.29 &     28.97 &    23.618 &     0.035 &    23.626 &     0.032 &    23.502 &     0.038 &        50 &      4.36 &      2.04 &      1.18 &     -52.9 &    24.600 &    21.018 &     30.46 &      3.41 &      0.02 &     0.342 &      0.72 &         0 &     1.270 &     0.085  \cr 
        19 &    733.06 &     33.22 &    24.383 &     0.062 &    24.448 &     0.052 &    24.377 &     0.068 &        32 &      4.20 &      1.84 &      1.00 &     -80.0 &    24.600 &    21.908 &     30.34 &      4.84 &      0.02 &     0.222 &      0.65 &         0 &     0.752 &     0.126  \cr 
        20 &    909.79 &     34.29 &    21.844 &     0.011 &    21.767 &     0.012 &    21.772 &     0.012 &       115 &      3.50 &      2.25 &      1.54 &     -74.9 &    24.600 &    19.339 &     30.16 &      2.70 &      0.02 &     0.532 &      0.83 &         0 &     0.914 &     0.021  \cr 
      .   &     &    .  &     &     . &     &     &     &     . &        &       &       &      . &      &     &     &     . &      &       &     .&      &          &     . &       \cr 
      .   &     &   .   &     &     . &     &     &     &     . &        &       &       &      . &      &     &     &     . &      &       &     .&      &          &     . &       \cr 
      .   &     &   .   &     &     . &     &     &     &     . &        &       &       &      . &      &     &     &     . &      &       &     .&      &          &     . &       \cr 
\noalign{\smallskip}
\noalign{\hrule}
\noalign{\smallskip}
\end{tabular}
%
%\end{center}
}
\end{sidetable}

\cleardoublepage
\singlespace
%\begin{center}
\centerline{\sc \hfil Table 4A\footnote{Complete versions of
Table 4 (Tables 4A--4K) are available on AAS CD-ROM series, Vol. ZZZ}
 \hfil }

\centerline{\sc \hfil Sample Morphology Catalog for A370\#2 \hfil }

{\scriptsize
\hspace{-1truein}\begin{tabular}{rccclrrcl}
\noalign{\medskip}
\hline\hline
\noalign{\smallskip}
{ID} & {DG \#} &  $I_{814}$ & $A$ & {Class} & {T-type} & {D} & {Int} & {Comments} \cr
\hline
\noalign{\smallskip}
   4 & ... & 22.123 &  0.052 &   Sc     &    5 &   0  & ... &   highly inclined disk system; arms, structure visible? \cr      
   7 & ... & 20.121 &  9.999 &   $\ast$?     &  $-$99 & $-$99  & ... &   frame edge -- star or bright object with jet \cr                
  10 & ... & 25.230 &  0.087 &   ?      &  $-$99 & $-$99  & ... &   cosmic ray clump? \cr                                           
  16 & ... & 22.330 &  0.088 &   Sd     &    7 &   2  &  T  &   late type, with fan like extension toward frame edge \cr        
  17 & ... & 21.966 &  0.075 &   Sc     &    5 &   1  & ... &   slightly wedge-shaped late type system \cr                      
  20 & ... & 21.772 &  0.078 &   E/S0   &   $-$4 & $-$99  & ... &   small E or S0 -- asymmetric and/or disturbed? \cr               
  23 & ... & 22.157 &  0.104 &   E      &   $-$5 &   0  & ... &   small, highly concentrated spheroid or star \cr                 
  27 & ... & 22.236 &  0.183 &   Sbc    &    4 &   2  & ... &   mild warp, near edge-on disk and bulge; no perturber \cr        
  30 & ... & 22.373 &  0.104 &   Sc     &    5 &   0  & ... &   edge-on late type; low SB \cr                                   
  36 & ... & 22.904 &  0.185 &   Sd/Irr &    8 &   2  &  I? &   late-type perturbed by nearby, concentrated dwarf \cr          
  38 & ... & 22.177 &  0.048 &   Sb/E?  &    3 &   2  &  T? &   spiral w/ disturb disk, or E w/ tidal tails? \cr                
  44 & ... & 20.667 &  0.080 &   Sa     &    1 &   0  & ... &   two smooth arm spiral with bulge; face on \cr                   
  46 & ... & 22.785 &  0.162 &   Sc     &    5 &   2  & ... &   small ring-like galaxy, chopped in two? \cr                     
  49 & ... & 22.208 &  0.065 &   E/S0   &   $-$4 &   0  & ... &   diffuse spheroid, or face on disk \cr                           
  54 & ... & 22.105 &  0.055 &   E      &   $-$5 &   0  & ... &   spheroid (distant?) with many surrounding companions \cr        
  67 & ... & 22.288 &  0.101 &   S0/a   &    0 &   1  & ... &   S0 or spiral with asymetric disk \cr                            
  73 & ... & 22.185 &  0.062 &   $\ast$      &  $-$99 & $-$99  & ... &    \cr                                                            
  75 & ... & 20.833 &  0.258 &   Scd    &    6 &   3  & M/I &   2 tidal tails from merger, or interr w/ compan @11? \cr         
  88 & ... & 21.910 &  0.096 &   Sd/Sm  &    7 &   1  & ... &   small, blobby disk \cr                                          
  89 & ... & 22.838 &  0.115 &   Sb     &    3 &   2  &  T? &   tiny disk system, prominent bulge; spiral or tidal arm \cr      
 .  &  & . &   &   .     &     &   .  &  &   . \cr      
 .  &  & . &   &   .     &     &   .  &  &   . \cr      
 .  &  & . &   &   .     &     &   .  &  &   . \cr      
\noalign{\smallskip}
\noalign{\hrule}
\noalign{\smallskip}
\end{tabular}
}
%
%\end{center}

\begin{thebibliography}{}
\itemsep=0in

\bibitem[Abrahams et al.\ 1995]{ra95}
Abrahams, R.G.,  Valdes, F., Yee, H.K.C.\ \& Van den Bergh, S.,
1995, ApJ, 432, 75.

\bibitem[Andreon et al.\ 1996]{and96}
Andreon, S., Davoust, E., Michard, R., Nieto, J.-L.\ \& Poulain, P., 1996,
A\&AS, 116, 429.

\bibitem[Barger et al.\ 1996]{ajb96}
Barger, A.J., Arag\'on-Salamnaca, A., Smail, I., Ellis, R.S.,
Couch, W.J., Dressler, A., Oemler, A., Butcher, H.\ \& Sharples, R.M.,
1996, in prep.

\bibitem[Bertin \& Arnouts 1996]{eb95} 
Bertin E.\ \& Arnouts, S., 1996, A\&A, in press.

\bibitem[Broadhurst, Ellis \& Shanks 1988]{bes88}
Broadhurst, T.J., Ellis, R.S.\ \& Shanks, T., 1988, MNRAS, 235, 827.

\bibitem[Butcher \& Oemler 1978]{bo78}
Butcher, H.\ \& Oemler, A., 1978, ApJ, 279, 18.

\bibitem[Butcher \& Oemler 1984]{bo84}
Butcher, H.\ \& Oemler, A., 1984, ApJ, 285, 426.

\bibitem[Castander et al.\ 1994]{fjc} Castander, F.J., 
Ellis, R.S., Frenk, C.S., Dressler, A.\ \& Gunn, J.E., 1994, ApJ, 424, L79.

\bibitem[Colless 1989]{col89}
Colless, M.M., 1989, MNRAS, 237, 799.

\bibitem[Couch \& Sharples 1987]{cs87}
Couch, W.J.\ \& Sharples, R.M., 
1987, MNRAS, 229, 423.

\bibitem[Couch et al.\ 1994]{cess94}
Couch, W.J., Ellis, R.S.,  Sharples, R.M.\ \&  Smail, I., 
1994, ApJ, 430, 121.

\bibitem[Cowie et al.\ 1996]{cow96}
Cowie, L.L., Songaila, A.\ \& Hu, E.M., 
1996, preprint.

\bibitem[de Vaucouleurs, de Vaucouleurs \& Corwin 1976]{dvdvc76}
de Vaucouleurs, Z., de Vaucouleurs, Z.\ \& Corwin, H., 1976.

\bibitem[Dressler 1980]{d80}
Dressler, A., 1980, ApJS, 42, 565.

\bibitem[Dressler \& Gunn 1992]{dg92}
Dressler, A.\ \& Gunn, J.E., 1992, ApJS, 78, 1.

\bibitem[Dressler et al.\ 1994]{dobg94}
Dressler, A., Oemler, A., Butcher, H.\  \& Gunn, J.E., 1994, ApJ, 430, 107.

\bibitem[Dressler et al.\ 1996]{ad96}
Dressler, A., Oemler, A., Smail, I., Couch, W.J., Ellis, R.S., Barger, A.,
Butcher, H., Poggianti, B.M.\ \& Sharples, R.M., 1996,
ApJ, submitted.

\bibitem[Ellis et al.\ 1996]{rse96}
Ellis, R.S., Smail, I., Dressler, A.,
Couch, W.J.,  Oemler, A., Butcher, H.\ \& Sharples, R.M.,
1996, ApJ, submitted.

\bibitem[Griffiths et al.\ 1994]{grif94}
Griffiths, R.E., Casertano, S., Ratnatunga, K.U., Neuschaefer, L.W.,
Ellis, R.S., Gilmore, G.F., Glazebrook, K. Santiago, B., Huchra, J.P.,
et al., 1994, ApJ, 435, L19.

\bibitem[Henry et al.\ 1982]{h82} 
Henry, J.P., Soltan, A., Briel, U.\ \& Gunn, J.E., 1982, ApJ, 262, 1.

\bibitem[Holtzman et al.\ 1995]{holt95} 
Holtzman, J.A., Burrows, C.J., Casterno, S., Hester, J.J., 
Trauger, J.T., Watson, A.M.\ \& Worthey, G., 1995,
PASP, 107, 1065.

\bibitem[Kodama \& Arimoto 1996]{ka96} Kodama, T.\ \&
Arimoto, N., 1996, preprint (astro-ph/9609160).

\bibitem[Kron 1980]{kron80} Kron, R.G., 1980, ApJS, 43, 305.

\bibitem[Lilly et al.\ 1995]{lil95}
Lilly, S.J., Tresse, L.,  Hammer, F., Crampton, D.\ \&
LeFevre, O., ApJ, 1995, 455, 108.

\bibitem[Oemler, Dressler \& Butcher 1996]{odb92}
Oemler, A., Dressler, A.\ \& Butcher, H., 1996, ApJ, submitted.

\bibitem[Sandage, Freeman \& Stokes 1970]{sfs70}
Sandage, A., Freeman, K.C.\ \& Stokes, R.A., 1970, ApJ, 160, 831. (SFS)

\bibitem[Soucail et al.\ 1988]{gs88}
Soucail, G., Mellier, Y., Fort, B.\ \& Cailloux, M., 1988, A\&AS, 73, 471.

\bibitem[Smail et al.\ 1996]{irs96}
Smail, I., Ellis, R.S.,
Dressler, A., Couch, W.J., Oemler, A., Butcher, H.\ \& Sharples, R.M.,
1996, ApJ, in press.

\bibitem[Wang \& Stocke 1993]{ws}
Wang, Q.D.\ \& Stocke, J.T., 1993, ApJ, 408, 71.

\bibitem[Worthey 1994]{w94}
Worthey, G., 1994, ApJS, 95, 107.

\end{thebibliography}
\end{document}